\pdfoutput=1
\documentclass[journal]{IEEEtran}
\usepackage{amssymb,stmaryrd,amsmath,amsfonts,rotating}
\usepackage[noadjust]{cite} \usepackage{color} 
\usepackage[vflt]{floatflt} \usepackage{epic}
\usepackage{color}

\newcommand{\x}{\underline{x}}

\newcommand{\y}{\underline{y}}

\newcommand{\erasure}{\text{?}}

\RequirePackage{bbm} 
\definecolor{TODO}{rgb}{0.6,0.6,0.6} 

\definecolor{TOCHECK}{rgb}{0.8,0.8,0.8} 

\newtheorem{theorem}{Theorem}
\newcommand{\btheo}{\begin{theorem}}
\newcommand{\etheo}{\end{theorem}}
\newcommand{\bproof}{\begin{proof}}
\newcommand{\eproof}{\end{proof}}
\newtheorem{definition}[theorem]{Definition}
\newcommand{\bdefi}{\begin{definition}}
\newcommand{\edefi}{\end{definition}}
\newtheorem{fact}[theorem]{Fact}
\newcommand{\bprop}{\begin{fact}}
\newcommand{\eprop}{\end{fact}}
\newtheorem{corollary}[theorem]{Corollary}
\newcommand{\bcor}{\begin{corollary}}
\newcommand{\ecor}{\end{corollary}}
\newtheorem{example}[theorem]{Example}
\newcommand{\bex}{\begin{example}}
\newcommand{\eex}{\end{example}}
\newtheorem{lemma}[theorem]{Lemma}
\newcommand{\blemma}{\begin{lemma}}
\newcommand{\elemma}{\end{lemma}}
\newtheorem{remark}[theorem]{Remark}
\newcommand{\bremark}{\begin{remark}}
\newcommand{\eremark}{\end{remark}}
\newtheorem{conj}[theorem]{Conjecture}
\newcommand{\bconj}{\begin{conj}}
\newcommand{\econj}{\end{conj}}

\newcommand{\naturals}{\ensuremath{\mathbb{N}}}
\newcommand{\integers}{\ensuremath{\mathbb{Z}}}

\def\0{{\tt 0}} 
\def\1{{\tt 1}} 
\def\?{{\tt *}} 
\newcommand{\BPsmall}{\ensuremath{\text{\tiny BP}}} 
\newcommand{\qed}{{\hfill \footnotesize $\blacksquare$}}

 \newcommand{\dr}{\mathtt{r}}
 \newcommand{\dra}{\mathtt{r}_1}
 \newcommand{\drb}{\mathtt{r}_2}
 \newcommand{\dl}{\mathtt{l}}
 \newcommand{\dla}{\mathtt{l}_1}
 \newcommand{\dlb}{\mathtt{l}_2}

\allowdisplaybreaks
\usepackage{graphicx}	
\begin{document} 
\title{Spatially Coupled Codes over the Multiple Access Channel}
\author{\authorblockN{ Shrinivas Kudekar\authorrefmark{1} and Kenta Kasai\authorrefmark{2} \\ }
\authorblockA{\authorrefmark{1} New Mexico Consortium  and CNLS, Los Alamos National Laboratory, New Mexico, USA\\
 Email: skudekar@lanl.gov} \\
\authorblockA{\authorrefmark{2} Dept. of Communications and Integrated Systems, Tokyo Institute of Technology, 152-8550 Tokyo, Japan.\\
Email: {kenta}@comm.ss.titech.ac.jp} \\
 }

\maketitle

\begin{abstract} We consider spatially coupled code ensembles over a multiple
access channel.  Convolutional LDPC ensembles are one instance of spatially
coupled codes. It was shown recently that, for transmission over the binary
erasure channel, this coupling of individual code ensembles has the effect of
increasing the belief propagation threshold of the coupled ensembles to the maximum a-posteriori
threshold of the underlying ensemble.  In this sense, spatially coupled codes
were shown to be capacity achieving. It was observed, empirically, that these
codes are universal in the sense that they achieve performance close to the
Shannon threshold for any general binary-input memoryless symmetric channels.
 
In this work we provide further evidence of the threshold saturation phenomena
when transmitting over a class of multiple access channel. We show, by density
evolution analysis and EXIT curves, that the belief propagation threshold of
the coupled ensembles is very close to the ultimate Shannon limit.  

\end{abstract}

\section{Introduction}
It has long been known that convolutional LDPC (or
spatially coupled) ensembles, introduced by Felstr{\"{o}}m and Zigangirov
\cite{FeZ99}, have excellent thresholds when transmitting over general
binary-input memoryless symmetric-output (BMS) channels. The fundamental reason
underlying this good performance was recently discussed in detail in
\cite{KRU10} for the case when transmission takes place over the binary erasure
channel (BEC). In the limit of large $L$ and $w$, the spatially-coupled LDPC
code ensemble $(\dl,\dr,L,w)$ \cite{KRU10} was shown to achieve the MAP
threshold of $(\dl,\dr)$ code ensemble (see last paragraph in this section for
the definition of the $(\dl, \dr, L, w)$ ensemble). This is the reason why they
call this phenomena {\em threshold saturation via spatial coupling}.  In a
recent paper \cite{LeF10}, Lentmaier and Fettweis independently formulated the
same statement as conjecture. 

The phenomena of threshold saturation seems not
to be restricted to the BEC. By computing EBP GEXIT curves
\cite{MMRU09}, it was observed in \cite{KMRU10} that threshold saturation also occurs for general BMS channels.  
In other words, in the limit of large $\dl$ (keeping $\frac{\dl}{\dr}$ constant), $L$ and $w$,  the
coupled ensemble $(\dl,\dr,L,w)$ achieves {\em universally} the capacity of the
BMS channels under belief propagation (BP) decoding.  Such universality is not a characteristic
feature of polar codes \cite{Ari09} and the irregular LDPC codes \cite{RSU01}.
According to the channel, polar codes need selection of frozen bits
\cite{DBLP:journals/corr/abs-0901-2207} and irregular LDPC codes need
optimization of degree distributions. 

The principle which underlies the good performance of spatially coupled
ensembles has been shown to apply to many other problems in communications, and
more generally computer science. To mention just a few, the threshold
saturation effect (dynamical threshold of the system being equal to the static
or condensation threshold) of coupled graphical models has recently been shown
to occur for compressed sensing \cite{KP10}, and a variety of graphical models
in statistical physics and computer science like the random $K$-SAT problem,
random graph coloring, and the Curie-Weiss model \cite{HMU10}.  Other
communication scenarios where the spatially coupled codes have found immediate
application is to achieve the whole rate-equivocation region of the BEC wiretap
channel \cite{RUAS10}, and to achieve the symmetric information rate for a
class of channels with memory \cite{KuKa11}. 

It is tempting to conjecture that the same phenomenon occurs for transmission
over general multi-user channels. We provide some empirical evidence via
density evolution (DE) analysis that this is indeed the case. In particular, we
compute EXIT curves for transmission over a multiple access channel (MAC) with
erasures.  We show that these curves behave in an identical fashion to the
curves when transmitting over the BEC. We compute fixed points (FPs) of the
coupled DE and show that these FPs have properties identical to the BEC case.

For a review on the literature on convolutional LDPC ensembles, we refer the
reader to \cite{KRU10} and the references therein.  As discussed in
\cite{KRU10}, there are many basic variants of coupled ensembles.  For the sake
of convenience of the reader, we quickly review the ensemble $(\dl, \dr, L,
w)$. This is the ensemble we use throughout the paper as it is the simplest to
analyze.

\subsection{$(\dl, \dr, L, w)$ Ensemble \cite{KRU10}}

We assume that the variable nodes are at sections $[-L, L]$, $L
\in \naturals$. At each section there are $M$ variable nodes, $M
\in \naturals$. Conceptually we think of the check nodes to be
located at all integer positions from $[- \infty, \infty]$.  Only
some of these positions actually interact with the variable nodes.
At each position there are $\frac{\dl}{\dr} M$ check nodes. It
remains to describe how the connections are chosen.  We assume that
each of the $\dl$ connections of a variable node at position $i$
is uniformly and independently chosen from the range $[i, \dots,
i+w-1]$, where $w$ is a ``smoothing'' parameter. In the same way,
we assume that each of the $\dr$ connections of a check node at
position $i$ is independently chosen from the range $[i-w+1, \dots,
i]$.  The design rate of the ensemble $(\dl, \dr, L, w)$, with $w \leq 2 L$,
is given by
$$R(\dl, \dr, L, w)  = 
(1-\frac{\dl}{\dr}) - \frac{\dl}{\dr} \frac{w+1-2\sum_{i=0}^{w} 
\bigl(\frac{i}{w}\bigr)^{\dr}}{2 L+1}.$$
A discussion on the above ensemble  can be found in \cite{KRU10}.


\section{Channel Model, Achievable Rate Region, Iterative Decoding and Factor Graph} 
\subsection{Binary Adder Channel with Erasures}
We consider the simplest synchronous 2-user multiple access channel, the binary adder channel (BAC) with erasure. 
More precisely, the inputs to the MAC are binary $X_1, X_2\in \{0,1\}$. The
users take on the values $0,1$ with equal probability. The subscripts 1,2 denote the two users. 
The output $Y\in\{0,1,2,\erasure\}$ is given by  
\begin{align*}
 Y&=
\left\{\begin{array}{ll}
 Z=X_1 + X_2  & \text{ with probability }1-\epsilon\\
\erasure& \text{ with probability }\epsilon,
\end{array}\right.
\end{align*}
where $\epsilon$ is the fraction of erasures. 

\subsection{Achievable Rate Region} 
 We assume that the two users do not coordinate their transmission. This implies
 that the joint input distribution has a product form. 
 Let $R_1$ and $R_2$ denote the transmission rates of the two users. 
  The achievable rate region is given as follows. 
\begin{align*}
 R_1\leq &I(X_1;Y \vert X_2), \\
 R_2\leq &I(X_2;Y \vert X_1), \\
 R_1+R_2\leq &I(X_1,X_2;Y).
\end{align*}
 The mutual information values above can be computed as  
\begin{align*}
 I(X_1;Y\vert X_2) & = I(X_2;Y \vert X_1)=1-\epsilon, \\
 I(X_1,X_2;Y)  & =\frac{3(1-\epsilon)}{2},\\
 I(X_1; Y) & = I(X_2; Y) =\frac{1-\epsilon}{2}.
\end{align*}
The Shannon limit is defined as the ultimate erasure threshold below which both users can
successfully decode using any decoder. Thus, the Shannon threshold is given by,   
\begin{align}\label{eq:shannonthreshold}
 \epsilon_{\mathrm{Sh}}=\min(1-R_1,1-R_2,1-\frac{2}{3}(R_1+R_2)).
\end{align}

\subsection{Factor Graph and Iterative Decoding}
\begin{figure}[htp]
\begin{centering}
\input{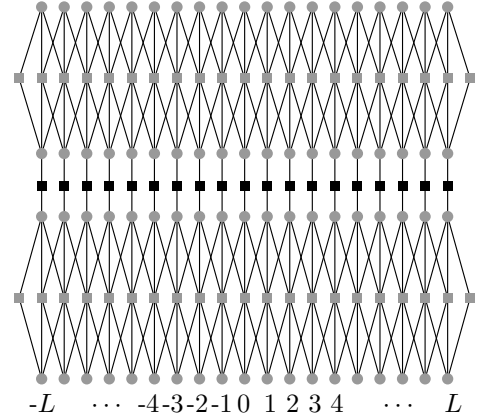}
\caption{The figure shows two protograph-based spatially coupled codes (each
belonging to one user) in light gray. The two protographs are connected by the
MAC function node shown in dark. Note that in the actual code the MAC function
node connects each variable node of one user to the corresponding variable node
of the other user. For the ease of illustration, we just show connections across 
one-half of the variable nodes. 
\label{fig:macfactorgraph}}
\end{centering}
\end{figure}

Figure~\ref{fig:macfactorgraph} shows the factor graph representation used in
the BP decoder analysis. The channel output is the vector $\y$ and the user inputs are
$\x_1$ and $\x_2$.  Each user has its own code and there is a function node
which connects the two factor graphs (dark squares in
Figure~\ref{fig:macfactorgraph}). This function node represents the channel
factor node $p(y_i \vert x_{1,i},x_{2,i})$ and we call it the MAC function node
(see \cite{RiU08, ADU02} for details).  Figure~\ref{fig:macfactorgraph} shows the spatially
coupled ensemble used by each user.  For the ease of illustration, we show the
protograph-based variant of spatially coupled codes.  If we do not use coupled
codes for transmission, then the two protographs above will be replaced by the
usual LDPC codes. 

The BP decoder passes messages between the various nodes in the factor graph.
The message passing schedule involves first passing the channel observations
from the MAC function nodes to the variable nodes of both of the users, then
performing one round of BP for both the users (in parallel) and then sending
the extrinsic information back to the MAC function node (from both the users).


\section{Uncoupled System: Density Evolution, Exit-like Curves}
\subsection{Density Evolution} Before we proceed to the analysis of coupled
codes, it is instructive to consider the DE analysis for the uncoupled $(\dl,
\dr)$-regular ensemble. More precisely, users 1 and 2 pick a code from the
 ensemble $(\dla, \dra)$-regular and $(\dlb, \drb)$-regular respectively.  From the schedule given above it is not hard to see
that for finite number of iterations and large blocklengths, the local
neighborhood around any node is a tree with high probability.  See \cite{RiU08,ADU02} for more details on
the DE setup.  Also, the BAC with erasures can be thought of as a BEC (for either user) with
erasure probability equal to $\epsilon + (1 - \epsilon)\mu/2$, where $\mu$ is
the erasure message flowing into the MAC function node. Indeed, the channel
output is either erased (wp $\epsilon$) or it is not erased (wp $1-\epsilon$) and we are still
uncertain of the transmitted symbol if the output is equal to 1 (occurs wp
1/2) and the other symbol is uncertain (wp $\mu$).  The FPs of the DE are then given by,
\begin{align*}
 y_{1}&=1-(1-x_{1})^{\dra-1}, \\
 x_{1}&=(\epsilon + \frac{1-\epsilon}{2}y_{2}^{\dlb})y_{1}^{\dla-1},  \\
 y_{2}&=1-(1-x_{2})^{\drb-1}, \\
 x_{2}&=(\epsilon + \frac{1-\epsilon}{2}y_{1}^{\dla})y_{2}^{\dlb-1},
\end{align*}
where $x_1 (y_1)$ and $x_2 (y_2)$ are variable-to-check (check-to-variable) erasure messages of user 1
and 2 respectively. Note that if $\dla=\dlb=\dl$ and $\dra=\drb=\dr$, then the above
equations reduce to a single parameter equation and is given by 
\begin{align*}
 &x = (\epsilon + \frac{(1-\epsilon)}2 y^{\dl})y^{\dl-1}, \\
 &y = 1 - (1-x)^{\dr-1}.
\end{align*}

\subsection{Exit-like Curves}
We define the BP EXIT-like\footnote{The reason we call this function EXIT-like
is because we do not provide any operational interpretation of these curves like
the Area theorem \cite{RiU08}. The curves are drawn only to illustrate that the
BP performance of coupled codes is close to the Shannon threshold, which is the main result of the paper.} function as follows. 
\begin{align}\label{eq:EBP}
h^{\mathrm{BP}}(\epsilon)= \frac{3}{2}y_{1}^{\dla}y_{2}^{\dlb} + y_{1}^{\dla}(1-y_{2}^{\dlb}) + (1-y_{1}^{\dla})y_{2}^{\dlb}.
\end{align}
An intuitive reason as to why we define the BP EXIT function as above is since
the entropy of $Z_i=X_{1,i}+X_{2,i}$ is $H(1/4,1/4,1/2)=3/2$ when a priori messages from both LDPC codes are erased 
 and since the entropy of $Z_i$ is 1 when either of them is erased and the other is not. 

Assume that all the FPs  are parametrized with $x_1$ such as $(x_1,y_1(x_1),x_2(x_1),y_2(x_1),\epsilon(x_1)). $
This assumption is true if $(\dla,\dra)=(\dlb,\drb)=(\dl,\dr)$ with 
\begin{align}
 y_{1}(x_1)&=y_2(x_1)=1-(1-x_{1})^{\dr-1}, \nonumber \\
 x_{2}(x_1)&=x_1, \nonumber \\
 \epsilon(x_1)&=\frac{\frac{x_1}{y_1(x_1)^{\dl-1}}-\frac{y_2(x_1)^{\dl}}{2}}{1-\frac{y_2(x_1)^{\dl}}{2}}.\nonumber
\end{align}
 We then have BP EXIT function as follows
\begin{align*}
h^{\mathrm{BP}}(x_1)=& \frac{3}{2}y_{1}(x_1)^{\dla}y_{2}(x_1)^{\dlb}\\
 &+ y_{1}(x_1)^{\dla}(1-y_{2}(x_1)^{\dlb}) \\
&+ (1-y_{1}(x_1)^{\dla})y_{2}(x_1)^{\dlb}.
\end{align*}
We also consider the extended BP EXIT (EBP EXIT) curve which is the plot of all
the fixed points of DE. In the case of codes (of each user) being picked from
the same ensemble, the EBP EXIT-like curve is given by the parametric curve $(h^{\BPsmall}(x), \epsilon(x))$,
 where $x$ is the variable-to-check node message of either code. 

\begin{example}\label{eg:MAPthreshold}
Figure~\ref{fig:EBP} shows the plot of the EBP EXIT-like curve for $\dla=\dlb=3,
\dra=\drb=6$. We choose this particular example since as seen from above it is
easier to evaluate the value of $\epsilon$ given a fixed value of $x$ (the
variable-to-check node message in either of the code). The BP threshold is
$\approx 0.12256$ which is much less than the Shannon threshold of $1/3$. 
\begin{figure}[htp]
\begin{centering}
\input{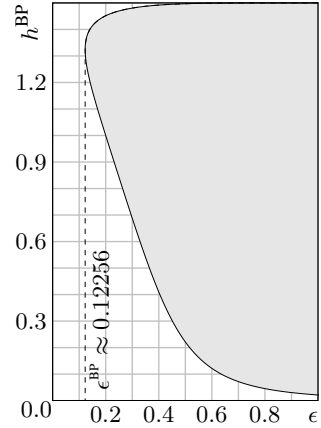}
\caption{EBP EXIT curve for the case when both users pick a code from the
$(3,6)$ and $(3,6)$. The BP threshold is $\approx 0.12256$ and the Shannon threshold is $1/3$. 
\label{fig:EBP}}
\end{centering}
\end{figure}
We observe that if we increase the degrees to $(4,8)$ for both
the codes, the BP threshold dramatically drops to zero. Also note the C
shape of the EXIT curve, indicating that there are exactly 3 FPs including a trivial FP plotted at $(0,\epsilon)$ for each
channel value, similar to the BEC case.  
\end{example}

\section{Main Results} In this section, we analyze the performance of  coupled
codes over BAC with erasures. We use the $(\dla, \dra, L, w)$ coupled ensemble
for user 1 and $(\dlb, \drb, L, w)$ ensemble for user 2. As a shorthand
notation we use $(\dla, \dra, \dlb, \drb, L, w)$ to denote both the ensembles.
Our main result is that, via DE analysis, the BP threshold of the coupled
ensemble is very close to the Shannon threshold given by
\eqref{eq:shannonthreshold}.  Furthermore, by increasing the degrees, the BP
threshold of the coupled ensemble goes to the Shannon threshold. 

Next, we 
develop the DE equation when transmitting using the coupled codes.  

\subsection{Density Evolution for the $(\dla,\dra,\dlb,\drb,L,w)$ ensemble} We develop the
DE equations assuming that the two users use ensembles of different degrees.
Consider the $(\dla, \dra, \dlb, \drb, L, w)$ ensemble.   To perform the DE
analysis, we  already take the limit $M\to \infty$ (the number of variable
nodes in each section).  

Let $x_{1,i}$, $i\in \integers$, denote the average erasure probability which
is emitted by variable nodes at position $i$ to check nodes at position $i$ for
user 1. Similarly define $x_{2,i}$ for the user 2. For $i \not \in [-L, L]$,
we set $x_{1,i}=x_{2,i}=0$.
For $i \in [-L, L]$ the DE is given by
\begin{align}\label{eq:densevolxi}
 y_{1,i}&=1-(1-\frac{1}{w}\sum_{k=0}^{w-1}x_{1,i-k})^{\dra-1},
 \nonumber \\
 x_{1,i}&=\big(\epsilon + \frac{1-\epsilon}{2}\big(\frac1w\sum_{j=0}^{w-1}y_{2,i+j}\big)^{\dlb}\big)\big(\frac1w\sum_{j=0}^{w-1}y_{1,i+j}\big)^{\dla-1},
 \nonumber \\
 y_{2,i}&=1-(1-\frac{1}{w}\sum_{k=0}^{w-1}x_{2,i-k})^{\drb-1},\nonumber \\
 x_{2,i}&=\big(\epsilon + \frac{1-\epsilon}{2}\big(\frac1w\sum_{j=0}^{w-1}y_{1,i+j}\big)^{\dla}\big)\big(\frac1w\sum_{j=0}^{w-1}y_{2,i+j}\big)^{\dlb-1}.
\end{align}
 We will use the notation $\epsilon^{\BPsmall}(\dla, \dra, \dlb, \drb, L, w)$ to denote the
 threshold of the BP decoder when we use coupled codes for transmission. Also, we use $\epsilon^{\BPsmall}(\dla, \dra,
 \dlb, \drb)$ to denote the BP threshold of the underlying uncoupled
 ensemble.

  As a shorthand we use
 $g_1(x^{1,2}_{i-w+1},\dots,x^{1,2}_{i+w-1})$ to denote
 $(\epsilon + \frac{1-\epsilon}{2}(\frac1w\sum_{j=0}^{w-1}y_{2,i+j})^{\dlb})(\frac1w\sum_{j=0}^{w-1}y_{1,i+j})^{\dla-1}$ and   
 also 
$g_2(x^{1,2}_{i-w+1},\dots,x^{1,2}_{i+w-1})$ to denote
 $(\epsilon + \frac{1-\epsilon}{2}(\frac1w\sum_{j=0}^{w-1}y_{1,i+j})^{\dla})(\frac1w\sum_{j=0}^{w-1}y_{2,i+j})^{\dlb-1}$. 

\begin{definition}[FPs of Density Evolution]\label{def:fixedpoints}Consider DE for the $(\dla, \dra, \dlb, \drb, L, w)$ ensemble.
Let $\x_1=(x_{1,-L}, \dots, {x}_{1,L})$ and $\x_2=(x_{2,-L}, \dots, {x}_{2,L})$ denote the vector of variable-to-check erasure messages for user 1 and 2 respectively. We call $\x_1$ and $\x_2$ the {\em constellation} of user 1 and 2 respectively. We say that $(\x_1, \x_2)$ forms a FP
of DE with channel $\epsilon$ if $(\x_1,\x_2)$ fulfills (\ref{eq:densevolxi})
for $i \in [-L, L]$.  As a shorthand we then say that $(\epsilon, \x_1, \x_2)$
is a FP.  We say that $(\epsilon, \x_1, \x_2)$ is a {\em non-trivial}
FP if either $\x_1$ or $\x_2$ is not identically equal to $0\,\,\forall\, i$.
Again, for $i\notin [-L,L]$, $x_{1,i}=x_{2,i} = 0$.
\qed
\end{definition}

\begin{definition}[Forward DE and Admissible Schedules]\label{def:forwardDE} 
Consider {\em forward} DE for the $(\dla, \dra, \dlb, \drb, L, w)$ ensemble.  More
precisely, pick a channel $\epsilon$ and initialize 
$\x^{(0)}_1=\x^{(0)}_2=(1, \dots, 1)$. Let $\x^{(\ell)}_1$ and $\x^{(\ell)}_2$ be the result of
$\ell$ rounds of DE for user 1 and 2 respectively. More precisely, $\x^{(\ell+1)}_1$ and $\x^{(\ell+1)}_2$ are generated from
$\x^{(\ell)}_1$ and $\x^{(\ell)}_2$ by applying the DE equation \eqref{eq:densevolxi} to each
section $i\in [-L,L]$,
\begin{align*}
x_{1,i}^{(\ell+1)} & = g_1(x_{i-w+1}^{1,2,(\ell)},\dots,x_{i+w-1}^{1,2,(\ell)}), \\
x_{2,i}^{(\ell+1)} & = g_2(x_{i-w+1}^{1,2,(\ell)},\dots,x_{i+w-1}^{1,2,(\ell)}),
\end{align*}
where we use the notation $x_{i}^{1,2,(\ell)}$ to denote  $(x_{1,i}^{(\ell)}, x_{2,i}^{(\ell)})$. 
We call this the {\em parallel} schedule.

More generally, consider a schedule in which in each step $\ell$
an arbitrary subset of the sections is updated, constrained only by
the fact that every section is updated in infinitely many steps. We
call such a schedule {\em admissible}. Again, we call $\x^{(\ell)}_1$ and $\x^{(\ell)}_2$
the resulting sequence of constellations.  \qed 
\end{definition}
One can show that if we perform forward DE under any admissible schedule, then
the constellations $\x^{(\ell)}_1$ and $\x_2^{(\ell)}$ converge to a FP of DE and this FP is
independent of schedule. This statement can be proved similar to the one in
\cite{KRU10, RiU08}. 

For the case when $\dla=\dlb$ and $\dra=\drb$ we have that for any FP,
$x_{1,i}=x_{2,i}$ and $y_{1,i}=y_{2,i}$ for all $i$.

\subsection{Forward DE -- Simulation Results} In the examples below, the Shannon
threshold is computed using equation \eqref{eq:shannonthreshold}.
\begin{example}[Equal Degrees -- BP goes to Shannon] We consider forward DE for
the coupled ensembles. More precisely, we fix an $\epsilon$ and initialize all
$x_{1,i}$ and $x_{2,i}$ to 1, for $i \in [-L,L]$.  Then we run the DE given by
\eqref{eq:densevolxi} till we reach a fixed-point.  We fix $L=200$. For
$\dla=\dlb=3$ and $\dra=\drb=6$, we have that $\epsilon^{\BPsmall}(3, 6, 3, 6,
200, 3) \approx 0.332287$. If we increase the degrees we get
$\epsilon^{\BPsmall}(4, 8, 4, 8, 200, 4) \approx 0.333195$,
$\epsilon^{\BPsmall}(5, 10, 5, 10, 200, 5) \approx 0.333286$. We observe that
 by increasing the degrees the BP threshold approaches the Shannon threshold of
$1/3$. On the other hand for the uncoupled codes, $\epsilon^{\BPsmall}(3,6,3,6)
\approx 0.12256$ and for larger degrees the BP threshold is zero.
\end{example}

\begin{example}[Unequal Degrees -- BP goes to Shannon]
We also consider the  more general case when the degrees are not equal. For
$\dla=5, \dra=10$ and $\dlb=6, \drb=13$ we get
$\epsilon^{\BPsmall}(5,10,6,13,500,10)\approx 0.307647$. The Shannon threshold in this
case is equal to $\approx 0.307692$. For $\dla=9, \dra=10$ and $\dlb=6,
\drb=10$ we get $\epsilon^{\BPsmall}(9,10,6,10,500,10)\approx 0.59992$ and the Shannon
threshold is $= 0.6$. 
\end{example}

\subsection{EXIT curve plots} 
\begin{figure}[htp]
\begin{centering}
\input{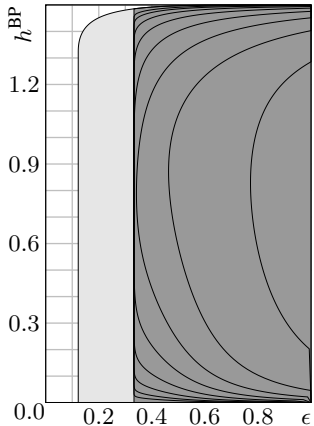}
\caption{ The EBP EXIT curve for $(3,6,3,6,L,3)$ with
$L=2,4,8,16,32,64,128,256$. The curve with light gray background is the BP EXIT
curve for the uncoupled $(3,6,3,6)$ ensemble.  We
see that as $L$ increases the EBP EXIT curves of the coupled system moves to the
left. The BP
threshold of the coupled system is $\approx 0.3323$ which is very close to the
Shannon threshold.   
\label{fig:EBPcoupled}}
\end{centering}
\end{figure}

We also show via EXIT analysis that the coupling of
regular LDPC codes pushes the BP threshold (of the coupled systems) to the Shannon threshold. 
For the purpose of illustration of the threshold saturation phenomena we focus only on the case
when $\dla=\dlb$ and $\dra=\drb$.  Thus, the variable-to-check node
messages, for any FP of DE, for both the users are equal (cf.
\eqref{eq:densevolxi}).  
Now, to plot the EBP EXIT curve, which is
 essentially the plot of all the fixed-points of DE, we define the entropy of  a
 constellation as 
\begin{align*}
\chi = \frac1{2L+1}\sum_{i=-L}^{L} x_{1,i}.
\end{align*}
To plot all the FPs of DE, we first fix a value of $\chi\in [0,1]$ and then run
the reverse DE process given in \cite{MMRU09}. Briefly, we start with an initial
variable-to-check message and run it through the check node. Then the
appropriate channel value is found such that the resulting constellation has
entropy equal to $\chi$. This process is run till we get an FP.
Figure~\ref{fig:EBPcoupled} shows the plot of the EBP EXIT curve for the 
 $(3,6,3,6,L,3)$ ensemble with $L=2,4,8,16,32,64,128,256$. We observe that the
 plot looks very similar to the case of single user transmission over a
 BEC. For small values of $L$ there is a
 large rateloss and the EBP EXIT curve is to the right. As $L$ increases, the
 rateloss diminishes and the curves move to the left. The limiting BP EXIT curve
 of the coupled system looks very similar to when we are transmitting over the
 BEC. It traces the BP EXIT function of the underlying uncoupled codes until the
 channel erasure value is very close to the Shannon threshold and then drops
 vertically to almost zero entropy. 

\subsection{Shape of the Constellation}
\begin{figure}[htp]
\begin{centering}
\input{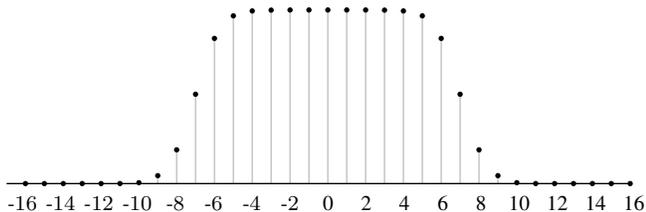}
\caption{ The unstable FP shown above has an entropy of $0.28$ and is obtained
via reverse DE. The constellation is symmetric around $0$ and is unimodal. 
 The flat middle
part has value close to $0.6548$ which is the value of stable FP for the
uncoupled system at $\epsilon \approx 0.3323$.    Both the users have identical FP constellation.
\label{fig:unimodalFP}}
\end{centering}
\end{figure}

Figure~\ref{fig:unimodalFP} shows the constellation of an unstable FP (which cannot be reached by BP).
This FP is obtained via the reverse DE process. This special FP  was the
key ingredient in proving threshold saturation over the BEC \cite{KRU10}.
Let us describe the (empirically observed) crucial properties of this
 constellation. 
 \begin{itemize}
 \item[(i)] The constellation is symmetric around $i=0$ and is unimodal. The
 constellation has $\epsilon\approx 0.3323$, which is close to the Shannon threshold of $1/3$. 
 \item[(ii)]  The value in the
 flat part in the middle is $\approx 0.6548$ which is very close to the stable
 FP of DE for the underlying uncoupled $(3,6)$-regular ensemble at $\epsilon \approx 0.3323$.  
 \item[(iii)] The transition from values close to zero to values close to $0.6548$
 is very quick.
 \end{itemize}

\section{Discussion}
In this paper we show that, by using coupled regular LDPC codes when transmitting over the 2 user
BAC with erasures, the BP threshold can be made very close to the Shannon threshold.
In this sense, the coupled codes are threshold saturating. We demonstrate this
by plotting EXIT-like curves. The behavior of these curves is very similar to
when transmitting over the BEC. Even the shape of the constellation of an
unstable FP of DE is same as the BEC case. Thus we believe one should be able to
provide a proof of this phenomena on the lines of the BEC proof \cite{KRU10}. 

Another interesting question is to determine area theorems which will also
further show that the BP threshold of the coupled system goes to the MAP
threshold of the underlying uncoupled codes, when we consider finite degrees.
 To do this we would need to define an appropriate EXIT function. 

Lastly, it would be interesting to see if we can demonstrate the threshold
saturation phenomena to more general MAC channels, like the 2 user BAC with
additive Gaussian noise. 

\section{Acknowledgments}
SK acknowledges support of NMC via the NSF collaborative grant CCF-0829945
on ``Harnessing Statistical Physics for Computing and Communications.'' 
 SK would also like to thank R\"udiger Urbanke for his encouragement.

\bibliographystyle{IEEEtran} 
\bibliography{lth,lthpub,kasai}
\end{document}